# Basis for a vector space generated by Hamiltonian time paths in a complete time graph

Malay Dutta

Ex-Professor

Indian Institute of Information Technology Guwahati

and Anjana K. Mahanta

Department of Computer Science Gauhati University

## Abstract

In this paper we introduce the notion of a complete time graph $K_n^T$. We define time paths and Hamiltonian time paths in $K_n^T$. Each Hamiltonian time path (htp) is associated with some permutation $p = (p_1, p_2, \ldots, p_n)$ of $\{1,2,\ldots,n\}$. The characteristic function of this path forms a vector in the vector space of rational-valued functions on the set of edges $E(K_n^T)$ of $K_n^T$. $H(K_n^T)$ will denote the vector space generated by these functions. The main result in this paper is to show that the dimension of $H(K_n^T)$ is $n(n-1)(n-2) + 1$ for $n \geq 5$. We also give a $O(n^5)$ algorithm for the construction of a basis in $H(K_n^T)$.

1) **Introduction :** We will be concerned here with the possible paths that a tourist may take within n + 1 cities 0, 1, 2, … , n. Suppose the tourist starts at city 0 on day 0, travels through city $p_i$ on day i for $1 \le i \le n$ ($p_i \ne p_{i+1}$), $1 \le p_i \le n$ and finally returns to city 0 on day n + 1. We can formulate the path that the tourist takes to be a path in a special directed graph. Graphs and directed graphs are defined and studied in [3]. A complete time-graph $K_n^T$ is defined essentially to be a directed graph with vertices (i, t) for $1 \le i, t \le n$ and edges (i,j,t) from (i,t) to (j,t+1) for $1 \le i,j \le n$ (i≠j) and $1 \le t \le n - 1$. Two additional vertices (0, 0) and (0, n+1) are also introduced. (0,0) is a source vertex with outgoing edge (0, j, 0) to (j, 1) for $1 \le j \le n$. (0, n+1) is a destination vertex with incoming edge (i, 0, n) from (i, n) for $1 \le i \le n$. It is easy to check that $K_n^T$ has $n^2 + 2$ vertices and $n(n-1)^2 + 2n$ edges. This definition of $K_n^T$ is based on the notion of a time dependent traveling salesman problem [4]. A complete time-graph was already introduced in [5]. Our present definition is a slight modification of the definition given in [5]. This is a natural generalization of a complete directed graph on the vertices 0, 1, 2, …, n. Thus in $K_n^T$, the tourist starts at the vertex (0, 0), goes through the vertices ($p_t$, t) for $1 \le t \le n$ and finally reaches (0, n + 1). In her tour the tourist thus goes through the edges (0, $p_1$, 0) followed by ($p_t$, $p_{t+1}$, t) for $1 \le t \le n - 1$, and finally through the edge ($p_n$, 0, n). A path from (0, 0) to (0, n + 1) will be called a time-path. The tourist thus follows a time-path. If the tourist goes through each of the cities 1, 2, ... , n precisely once, then her path is called a Hamiltonian time-path (htp) when $(p_1, p_2, ... , p_n)$ is a permutation of $\{1, 2, ... , n\}$. In the associated complete directed graph the corresponding path is a Hamiltonian cycle.

In section 2, we give some preliminary results on finite dimensional vector spaces over the field of rational numbers. In section 3 we shall state some more definitions and also state the problem we solve in this paper. In sections 4 and 5 we will present our main results. In section 6 we give the conclusions and the directions of possible future work.

2) **Finite dimensional vector spaces over rational numbera:** In vector spaces over rational numbers, because of the absence of square roots we have to give up the notion of norms. But most the other notions including inner products, orthogonality etc are retained. Also the notion of annihilators and spaces of annihilators are introduced. Let V be a finite dimensional vector space over the field Q of rational numbers. Then V is isomorphic to $Q^X$, the space of rational valued functions on $X = \{1, 2, ... , n\}$, which is a vector space over Q under point-wise operations. If X is empty, $Q^X$ will taken to be the zero vector. It is then obvious that $\dim(Q^X)$, the dimension of $Q^X$ is $|X|$.

For $f, g \in Q^X$ we define an inner product $\langle f, g \rangle = \sum_{x \in X} f(x) g(x)$. Obviously $\langle f, g \rangle$ is symmetric and bilinear. Also $\langle f, f \rangle = 0$ iff $f = 0$. $f, g$ are called orthogonal if $\langle f, g \rangle = 0$. $f$ is said to annihilate $g$, or $f$ is said to be an annihilator of $g$, if $f, g$ are orthogonal. For a subset $S$ of $Q^X$, $f$ is said to annihilate $S$ if $f$ annihilates $g$ for every $g$ in $S$. The set of annihilators of $S$ is denoted by $A(S)$ and is obviously a subspace of $Q^X$. Given a sequence of linearly independent elements $f_1, f_2, \ldots, f_n$ of $Q^X$ a sequence of non-zero mutually orthogonal elements $g_1, g_2, \ldots, g_n$ can be formed such that $g_i$ is in $\langle f_1, f_2, \ldots f_i \rangle$ for $1 \leq i \leq n$. This is done through a process similar to Gram-Scmidt orthonormalization in Hilbert spaces discussed in [2]. The following theorem is easily proved. We will give here an outline of the proof. The theorem will have an important application in the work to be presented in the rest of this paper.

Theorem 2.1 : Let V be a subspace of $Q^X$. Then $\dim(V) + \dim(A(V)) = |X|$.

Proof : Let $\dim(V) = k$, $\dim(Q^X) = |X| = n$. Take a basis $f_1, f_2, \ldots, f_k$ of V and extend it to a basis $f_1, f_2, \ldots, f_k, f_{k+1}, \ldots, f_n$ of $Q^X$. Orthogonalize this obtaining $g_1, g_2, \ldots, g_n$ as discussed above. It is easy to show now that $g_1, g_2, \ldots, g_k$ is a basis for V and $g_{k+1}, \ldots, g_n$ is a basis for A(H). Hence $\dim(V) = k$ and $\dim(A(V)) = n - k$. Hence $\dim(V) + \dim(A(V)) = n = |X|$. This proves the result.

3)   The Problem solved in this paper :   Since the characteristic function of a htp h is a rational-valued function on $E(K_n^T)$ the set of edges of $K_n^T$, we can consider h to be a vector in the space Q-valued functions on $E(K_n^T)$, which forms a vector space over Q the set of rational numbers. A htp will usually be identified with this vector. A time-graph G of order n is defined to be a sub-graph of $K_n^T$ with the same vertices as that of $K_n^T$, and the set of edges E(G) a subset of $E(K_n^T)$. Let H(G) denote the vector space generated by the htp's in G i.e. those htp's whose edges lie in G. In section 4, we prove that $\dim(H(K_n^T)) \leq d_n = n(n-1)(n-2)+1$ for $n \geq 5$. Let $f_1, f_2, ..., f_n$ be any sequence of functions on $E(K_n^T)$ such that there is a sequence of edges $x_1, x_2, ..., x_n$ satisfying $f_i(x_i) = 1$ and $f_i(x_j) = 0$ if $i > j$. Then the sequence of functions is called an upper triangular sequence of functions and the sequence of edges is called the sequence of pivot edges. Clearly such a sequence of functions is linearly independent. In section 5, we give a $O(n^5)$ algorithm for constructing a number of $d_n$ upper triangular sequence of functions in $H(K_n^T)$ for $n \geq 5$. These functions are then linearly independent htps in $K_n^T$ for $n \geq 5$. Since this is the maximum number of linearly independent htp's in $K_n^T$, the constructed htp's will give the required basis of $H(K_n^T)$ for $n \geq 5$.

4) **Maximum dimension of $H(K_n^T)$**: In this section we prove that $\dim(H(K_n^T)) \leq d_n$ for $n \geq 5$. For $1 \leq i, t \leq n$, let $f_{(i,t)}$ be the characteristic function of a path from $(0, 0)$ to $(i, t)$. For $1 \leq i \leq n-1$, let $f_i$ be the characteristic function of a time-path that does not go through city $n$ at all, goes through city $i$ twice on non-consecutive days and goes through all other cities exactly once. For example for $1 < i < n-1$ (Note that since $n \geq 5$, value of $i$ can go upto 3) $f_i$ is 0 i 1 i 2 3 .. i-1 i+1 .. n-1 0, $f_1$ is 0 1 2 1 3 4 .. n-1 0 and $f_{n-1}$ is 0 n-1 1 n-1 2 3 .. n-2 0. We now define $e_{ijt}$ to be a function on E which is 1 on the edge $(i, j, t)$ and 0 on all other edges. For $1 \leq i, t \leq n$ define the functions $g_{(i,t)} = \sum_j e_{ijt} - \sum_j e_{jit-1}$ (In $e_{ijt}$, j=0 if t=n and in $e_{jit-1}$, j=0 if t=1). Also for $1 \leq i \leq n-1$ define $g_i = \sum_{jt} e_{ijt} - \sum_j e_{0j0}$. It is easy to verify that $\langle f_k, g_{(i,t)} \rangle = 0$ for $k = 1, 2, .., n-1$ and $1 \leq i, t \leq n$, $\langle f_{(i,t)}, g_{(i',t')} \rangle = -\delta_{i\,i'}\,\delta_{t\,t'}$ for $1 \leq i,t,i',t' \leq n$ and $\langle f_i, g_j \rangle = \delta_{ij}$ for $1 \leq i, j \leq n-1$. Let $\sum_{j=1,n-1} \alpha_j g_j + \sum_{i,t} \alpha_{(i,t)} g_{(i,t)} = 0$. Then taking inner product with $f_k$ for $k = 1, 2, ..n-1$, $\alpha_k = 0$, and taking inner product with $f_{(i',t')}$ for $1 \leq i', t' \leq n$, $\alpha_{(i',t')} = 0$. Hence $g_1, g_2, ...., g_{n-1}, \{g_{(i,t)}\}_{1 \leq i,t \leq n}$ are linearly independent. It is easy to see that for h in $H(K_n^T)$, $g_i$ for $1 \leq i \leq n-1$ and $g_{(i,t)}$ for $1 \leq i,t \leq n$ annihilate h. Thus these $n^2 + n - 1$ functions are linearly independent and are in $A(H_n)$. Thus $\dim(A(H_n)) \geq n^2 + n - 1$ for $n \geq 5$. By Theorem 2.1 $\dim(H(K_n^T)) + \dim(A(H(K_n^T))) = |E(K_n^T)| = 2n + (n-1)^2 n$. Hence $\dim(H_n) = 2n + (n-1)^2 n - \dim(A(H_n)) \leq 2n + (n-1)^2 n - n^2 - n + 1 = n(n-1)(n-2) + 1 = d_n$. In section 5, for $n \geq 5$, we shall inductively construct an upper-triangular basis for

$H(K_n^T)$ consisting of $d_n$ elements. This will prove that for $n \geq 5$,
(i) $\dim(H(K_n^T)) = d_n = n(n-1)(n-2)+1$, (ii) $\dim(A(H(K_n^T))) = n^2 + n - 1$
and (iii) $g_1, g_2, \ldots, g_{n-1}, \{g_{(i,t)}\}_{1 \leq i, t \leq n}$ forms a basis of $A(H(K_n^T))$.

5) Construction of a basis in $H(K_n^T)$ for $n \geq 5$ : The construction will be inductive. The start of the induction will be an upper triangular basis of $d_5 = 61$ elements in $H(K_5^T)$. This is given in the Appendix. The induction step will be to extend the upper-triangular basis of $d_{n-1}$ elements in $H(K_{n-1}^T)$ to an upper-triangular basis of $d_n$ elements in $H(K_n^T)$. For convenience we represent the htp's of $H(K_n^T)$ by the elements $p$ in $S_n$ the set of permutations of $\{1,2,\ldots,n\}$. By induction hypothesis an upper-triangular basis $\{q_i\}_{i=1}^{d_{n-1}}$ has been constructed in $H(K_{n-1}^T)$. We shall use this to construct an upper triangular set $\{p_i\}_{i=1}^{d_n}$ of htp's in $H(K_n^T)$. We first construct $\{p_i\}_{i=1}^{d_{n-1}}$ by (see Fig 1)

$p_i(k) = q_i(k)$ for $1 \leq k \leq n-1$, $1 \leq i \leq d_{n-1}$
$p_i(n) = n$ for $1 \leq i \leq d_{n-1}$
Clearly $[p_i\}_{i=1}^{d_{n-1}}$ is also upper triangular.

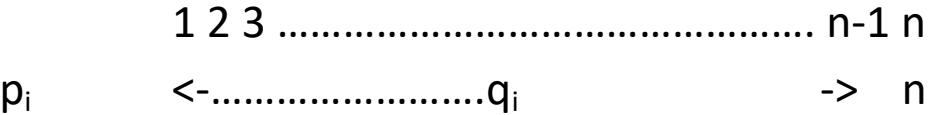

                1 2 3 ………………………………………………… n-1 n
  $p_i$        <-………………………$q_i$                     -> n

Fig 1

We now take a sequence of htps in $H(K_n^T)$ as $\{\{f_{ij}\}_{j=1}^{n-3}, g_i, h_i\}_{i=1}^{n-1}$ (see Fig 2) where

$f_{ij}(1) = g_i(1) = h_i(1) = n$

$f_{ij}(2) = g_i(2) = h_i(2) = i$

$f_{ij}(n-1) = g_i(n) = i \bmod (n-1) + 1$

$f_{ij}(n) = (i+j) \bmod (n-1) + 1$

$h_i(n-1) = g_i(n-1) = (i+1) \bmod (n-1) + 1$

$h_i(n) = (i+2) \bmod (n-1) + 1$

$f_{ij}$'s, $g_i$'s and $h_i$'s are defined arbitrarily at other points to form htp's. We now drop the last element $h_{n-1}$ from this set and insert the rest $(n-1)^2 - 1$ elements before the $p_i$'s to obtain the sequence $\{p_i\}_{i=1}^K$ where $K = d_{n-1} + (n-1)^2 - 1$ and it is easy to see that this sequence is also upper triangular.

| Layer | 1 | 2 | 3 | ... | n-2 | n-1 | n |
|---|---|---|---|---|---|---|---|
| $f_{11}$ | | n | 1 | < ............. Arbitrary to form htp .......> | | 2 | 3 |
| $f_{12}$ | | n | 1 | | | 2 | 4 |
| . | | | | | | | |
| . | | | | | | | |

| | | | | | |
|---|---|---|---|---|---|
| $f_{1\,n-3}$ | n | 1 | | 2 | n−1 |
| $g_1$ | n | 1 | | 3 | 2 |
| $h_1$ | n | 1 | | 3 | 4 |
| $f_{21}$ | n | 2 | | 3 | 4 |
| $f_{22}$ | n | 2 | | 3 | 5 |
| . | | | | | |
| . | | | | | |
| $f_{2\,n-3}$ | n | 2 | | 3 | 1 |
| $g_2$ | n | 2 | | 4 | 3 |
| $h_2$ | n | 2 | | 4 | 5 |
| . | | | | | |
| . | | | | | |
| $f_{n-2\,n-3}$ | n | n-2 | | n-1 | n-3 |
| $g_{n-2}$ | n | n-2 | | 1 | n-1 |
| $h_{n-2}$ | n | n-2 | | 1 | 2 |
| $f_{n-1\,1}$ | n | n-1 | | 1 | 2 |
| $f_{n-1\,2}$ | n | n-1 | | 1 | 3 |

.

.

| $f_{n-1\ n-3}$ | n | n-1 | | 1 | n-2 |
| $g_{n-1}$ | n | n-1 | | 2 | 1 |

Fig 2

Finally we take another sequence of upper triangular sequence of htp's $\{c_{ij}\}_{i=2}^{n-1}{}_{j=1}^{2n-3}$ (see Fig 3) where

$c_{ik}(i) = n$

$c_{ik}(i-1) = $ ceiling$(k/2)$

$c_{ik}(i+1) = $ ceiling$((k+1)/2)$ mod $(n-1) + 1$

$c_{ik}$'s are defined arbitrarily at other points to form htp's.

| | 1 | 2 | 3 | 4 | ………………… | i-1 | i | i+1 | ………………………… | n-2 | n-1 | n |
|---|---|---|---|---|---|---|---|---|---|---|---|---|
| $c_{21}$ | 1 | n | 2 | <……………….arbitrary to formhtp……………………………> |
| $c_{22}$ | 1 | n | | 3 | <……………………………………………………………………………………….> |
| $c_{23}$ | | 2 | n | 3 | <……………………………………………………………………………………….> |
| $c_{24}$ | | 2 | n | | 4 | <…………………………………………………………………………….> |

.

.

$c_{2\ 2n-5}$ n-2 n n-1 <……………………………………………………………………….>

$c_{2\ 2n-4}$ n-2 n 1 <……………………………………………………………………….>

$c_{2\ 2n-3}$ n-1 n 1 <……………………………………………………………………….>

$c_{31}$ <.. 1 n 2 ………………………………………………………………>

$c_{32}$ <.. 1 n 3 ………………………………………………………………>

.

.

$c_{3\ 2n-3}$ <.. n-1 n 1 …………………………………………………………….>

.

.

$c_{i1}$ <………………………………….1 n 2 ………………………………….>

$c_{i2}$ <………………………………….1 n 3………………………………….>

.

.

$c_{i\ 2n-3}$ <………………………………….n-1 n 1 ……………………………….>

.

.

$c_{n-1\ 1}$ <................................................................1  n  2>

$c_{n-1\ 2}$ <................................................................1  n  3>

.

.

$c_{n-1\ 2n-3}$ <...............................................................n-1  n  1>

Fig 3

We now insert these (n-2)(2n-3) elements above the previously constructed $p_i$'s to get the sequence $\{p_i\}_{i=1}^Q$ where $Q = d_{n-1}+(n-1)^2-1 +(n-2)(2n-3)$. It is easy to check that these Q elements are upper triangular and hence linearly independent. The sequence of pivot edges is given as follows. First comes the following sequence of pivot edges of (n-2)(2n-3) elements : $\{(n,2,i),(1,n,i-1),(n,3,i),(2,n,i-1), \ldots , (n,n-1,i),(n-2,n,i),(n,1,i)\}_{i=2}^{i=n-1}$. After that we get the sequence of pivot edges of $(n-1)^2-1$ elements :

{(2,3,n-1),(2,4,n-1), …, (2,n-1,n-1),(3,2,n-1),(n,1,1),

(3,4,n-1),(3,5,n-1), …, (3,1,n-1),(4,3,n-1),(n,2,1),

(4,5,n-1),(4,6,n-1), …, (4,2,n-1),(5,4,n-1),(n,3,1),

……

(n-1,1,n-1),(n-1,2,n-1), …, (n-1,n-3,n-1),(1,n-1,n-1),(n,n-2,1),

(1,2,n-1),(1,3,n-1), …, (1,n-2,n-1),(2,1,n-1)}.

Finally by the induction hypothesis, we get the sequence of pivot edges of the upper triangular basis of $H(K_{n-1}^T)$ consisting of $d_{n-1} = (n-1)(n-2)(n-3)+1$ elements. Note that the elements not specified in the htp's above can be easily filled to form htp's since n≥6 in the induction step. The case of n = 5 is covered in the induction basis, the proof of which appears in the Appendix through an explicit construction.

Adding up we see that $Q = n(n-1)(n-2) + 1 = d_n$. Since there are at most $d_n$ linearly independent elements in $H(K_n^T)$, these elements form an upper triangular basis in $H(K_n^T)$ and thus the dimension of $H(K_n^T)$ is n(n-1)n-2)+ 1 for n ≥ 5. After the construction of a basis in $H(K_{n-1}^T)$ the construction of a single $p_i$ in $H(K_n^T)$ needs O(n) time. There are $O(n^3)$ $p_i$'s to be constructed. So after the computation of the basis for $H(K_{n-1}^T)$ we need $O(n^4)$ additional steps to construct the basis for $H(K_n^T)$. Hence the complexity of the construction of a basis in $H(K_n^T)$ is $O(n^5)$.

6) **Conclusions and Future work :** In this paper we have given an $O(n^5)$ algorithm for the construction of a basis in $H(K_n^T)$ for n ≥ 5. We can try to generalize this to the construction of a basis in H(G) for an arbitrary time graph. Given a time graph G it will be called Hamiltonian if it has a htp. We can try a crude method to determine whether a given time graph G is Hamiltonian in the following way. We generate all the permutations and check if there is one giving a htp in G. Since there are exponentially large number of htps this will be exponential time. Since G will be Hamiltonian iff dim (H(G)) > 0 and since  bases in H(G) have number of elements bounded by a polynomial in n, taking the approach of computing a basis may be of help in evolving a polynomial-time algorithm.

# Appendix

Upper-triangular basis in $H(K_5^T)$ (The pivot edges are underlined)

| | | | | | |
|---|---|---|---|---|---|
| 1) | <u>1 5</u> 2 3 4 | 29) | <u>4 3</u> 1 2 5 | 57) | 5 2 4 <u>1 3</u> |
| 2) | 3 <u>5 2</u> 1 4 | 30) | <u>3 4</u> 1 2 5 | 58) | 5 <u>3 2</u> 4 1 |
| 3) | <u>3 5</u> 4 2 1 | 31) | 5 2 1 <u>4 3</u> | 59) | 5 2 3 <u>4 1</u> |
| 4) | 2 <u>5 4</u> 3 1 | 32) | 5 <u>2 1</u> 3 4 | 60) | 5 <u>2 4</u> 3 1 |
| 5) | <u>2 5</u> 1 3 4 | 33) | 5 4 <u>1 3</u> 2 | 61) | 5 4 2 <u>3 1</u> |
| 6) | 4 <u>5 1</u> 2 3 | 34) | 5 1 4 <u>3 2</u> | | |
| 7) | <u>4 5</u> 3 1 2 | 35) | 5 1 2 <u>3 4</u> | | |
| 8) | 3 <u>1 5</u> 2 4 | 36) | 5 3 <u>1 4</u> 2 | | |
| 9) | 1 3 <u>5 2</u> 4 | 37) | 5 1 3 <u>4 2</u> | | |
| 10) | 1 <u>3 5</u> 4 2 | 38) | 5 <u>3 1</u> 2 4 | | |
| 11) | 1 2 <u>5 4</u> 3 | 39) | 5 1 3 <u>2 4</u> | | |
| 12) | 3 <u>2 5</u> 1 4 | 40) | 5 4 <u>1 2</u> 3 | | |
| 13) | 3 4 <u>5 1</u> 2 | 41) | 5 1 4 <u>2 3</u> | | |
| 14) | 2 <u>4 5</u> 3 1 | 42) | 5 3 <u>4 2</u> 1 | | |
| 15) | 3 4 <u>1 5</u> 2 | 43) | 5 4 3 <u>2 1</u> | | |
| 16) | 1 4 3 <u>5 2</u> | 44) | 5 <u>3 4</u> 1 2 | | |
| 17) | 1 2 <u>3 5</u> 4 | 45) | 5 4 3 <u>1 2</u> | | |
| 18) | 1 3 2 <u>5 4</u> | 46) | 1 <u>4 3</u> 2 5 | | |
| 19) | 3 4 <u>2 5</u> 1 | 47) | 4 1 3 <u>2 5</u> | | |
| 20) | 3 2 4 <u>5 1</u> | 48) | <u>1 4</u> 2 3 5 | | |
| 21) | 2 1 <u>4 5</u> 3 | 49) | <u>4 1</u> 2 3 5 | | |
| 22) | 2 3 4 <u>1 5</u> | 50) | 2 <u>1 4</u> 3 5 | | |
| 23) | <u>1 3</u> 2 4 5 | 51) | 1 2 4 <u>3 5</u> | | |
| 24) | <u>3 1</u> 2 4 5 | 52) | <u>1 2</u> 3 4 5 | | |

25) <u>2 3</u> 1 4 5      53) <u>2 1</u> 3 4 5
26) <u>3 2</u> 1 4 5      54) 5 2 <u>3 1</u> 4
27) <u>2 4</u> 1 3 5      55) 5 3 2 <u>1 4</u>
28) <u>4 2</u> 1 3 5      56) 5 4 <u>2 1</u> 3